\documentclass[letterpaper]{article}
\usepackage{subcaption}

\usepackage{natbib,alifeconf}  
\usepackage{comment}
\usepackage{hyperref}
\usepackage[super]{nth}

\title{Two Ways of Understanding Social Dynamics: Analyzing the Predictability of Emergence of Objects in Reddit r/place Dependent on Locality in Space and Time}
\author{Alyssa M Adams$^{1,2,3}$, Javier Fernandez$^1$ \and Olaf Witkowski$^1$ \\
\mbox{}\\
$^1$Cross Labs, Kyoto, Japan \\
$^2$Morgridge Institute for Research, USA \\
$^3$Department of Bacteriology, University of Wisconsin-Madison, USA \\
alyssa.gp.adams@gmail.com} 

\begin{document}
\maketitle

\begin{abstract}
Lately, studying social dynamics in interacting agents has been boosted by the power of computer models, which bring the richness of qualitative work, while offering the precision, transparency, extensiveness, and replicability of statistical and mathematical approaches. A particular set of phenomena for the study of social dynamics is Web collaborative platforms. A dataset of interest is r/place, a collaborative social experiment held in 2017 on Reddit, which consisted of a shared online canvas of 1000 pixels by 1000 pixels co-edited by over a million recorded users over 72 hours. In this paper, we designed and compared two methods to analyze the dynamics of this experiment. Our first method consisted in approximating the set of 2D cellular-automata-like rules used to generate the canvas images and how these rules change over time. The second method consisted in a convolutional neural network (CNN) that learned an approximation to the generative rules in order to generate the complex outcomes of the canvas. Our results indicate varying context-size dependencies for the predictability of different objects in r/place in time and space. They also indicate a surprising peak in difficulty to statistically infer behavioral rules towards the middle of the social experiment, while user interactions did not drop until before the end. The combination of our two approaches, one rule-based and the other statistical CNN-based, shows the ability to highlight diverse aspects of analyzing social dynamics.
\end{abstract}

\section{Introduction}

The field of social dynamics studies behaviors that result from groups of interacting individuals that self-organize in particular ways. It is also one of the pillars of complexity science, and has ramifications in sociology, psychology, economics, animal behavior, and numerous fields. One of the most data-rich areas for the study of such social phenomena can be found in online communities, in particular on collaborative platforms such as wikis, Q\&A websites, and social media. This project focuses on Reddit, an online discussion platform that also hosted a collaborative social experiment on April Fools' Day of 2017, called Place (or r/place, the sub-community created for the occasion). The experiment involved an online canvas, which registered users could edit by changing the color of a single pixel from a 16-color palette. After each pixel was placed, a timer prevented the user from placing any pixels for a period of time between 5 and 20 minutes \citep{howWe2017}.


In just 72 hours, over a million registered Reddit users placed 16.5 million pixels to transform a simple, blank, 1000$\times$1000-pixel canvas into a surprisingly beautiful clash of communities, nations, and ideologies. Because each user could only place one pixel every 5-20 minutes, any single individual would have struggled to create a meaningful image on their own. However, through community collaboration, users quickly produced complex creations, surpassing all of our expectations about how this project would turn out once the 72 hours were up. Reddit released pixel-by-pixel placement data and additional community efforts were spurred to produce additional canvas analysis. The r/place Atlas project \citep{RPlaceAtlas2017} identified almost 1500 different objects and phenomena on the final state of the canvas, although does not identify objects and phenomena in intermediate canvas states.


Although most of the placement happened within a span of 72 hours, the full dataset from \cite{placedatasets2017} included pixels placed prior to the 72 hours. As a result, the full dataset includes pixel placements from 2017-03-21 21:03:09 UTC - 2017-04-03 16:58:55 UTC. In addition, some users were unable to place new pixels after 5 minutes and needed to wait up to 20 minutes (see posts such as \cite{20min}).

\section{Previous work and motivation}

In the current literature, there are some studies that have addressed the dynamics of this experiment. One example is \cite{litherland2021}, who studied the r/place event through the evolution of two types of objects, visual artifacts and social artifacts, that changed continuously over 72 hours. But because of the complex nature of the data, this study mainly focuses on a single image on the final canvas, along with its corresponding social community on the forums: The Mona Lisa replication. The Mona Lisa was present throughout most of the 72-hour span of r/place which makes it ideal to study its dynamics in comparison to post/comment activity that users used to coordinate pixel placement. 

The authors found interesting and complex interactions between the visual and social artifacts that supported the creation, stabilization, and preservation of the Mona Lisa image. These interactions are similar to top-down and bottom-up dynamics seen across several scales of biology \citep{walker}; bottom-up pixel placement of the initial image spurred social interactions, which lead to further image development and preservation from attacks in a top-down approach. On one hand, parts of the image (such as the face) initially appeared spontaneously, while on the other hand, the rest of the image was filled by a coordinated effort to complete and maintain the image against attacks. In context of emergence, the authors argue that the robustness of the Mona Lisa image is due to organized efforts on multiple levels, not only on a pixel-by-pixel basis, but also as a result of dynamics within the social artifact. These results suggest it may not be possible to understand the dynamics of emergent r/place structures based only on the pixel-by-pixel dataset. Instead, emergent structures are more fully understood using additional data, particularly on a social level as captured by subreddit posts, comments, and upvotes. In addition, \cite{rappaz2018latent} proposed a predictive method based on the graph of user interaction clustering that captures the latent structure of the emergent collaborative efforts, and showed that the method provides an interpretable representation of the social structure.

So far, these approaches rely on visual and social artifacts that are annotated externally and robust over time. Structures like the Mona Lisa image are static and do not change location on the canvas, however there are several documented objects that do not have well-defined borders, move location over time, or change shape and structure drastically over the course of their evolution. For example, ``The Void'' is an amorphous block spot that is defined as behavior that simply changes adjacent pixels from their current color to black. ``Rainbow Road'' is a rainbow path that spans a large area of the canvas and is moved throughout its evolution. In addition, objects such as national flags are documented to expand and collide with other objects. Objects and images compete for space, invade one another, and can spawn inside each other. Not all of these events are documented or have a corresponding social artifact to represent them.


In our analysis, we further generalize these approaches by developing a framework aimed at detecting emergent objects and artifacts over time \textit{without} relying on external annotations or social artifacts. Not all objects are given names, timestamps, and coordinate locations on the r/place canvas, therefore we are interested in a framework that can identify a wide variety of object types over the canvas evolution. We introduce a framework inspired by \cite{flack}, which demonstrates that individuals (like objects, images, and artifacts) can be defined in many different ways by constants throughout time or space. Within some time series data, particularly pixel-by-pixel placements in r/place, we generalize the concept of an ``individual'' image or object as a set of \textit{spatially local} rules that don't change over time.

However, we recognize the shortcomings of this approach. Because images that are reproduced in this canvas are defined externally, it is very unlikely that any image can be predicted before pixels are placed. In addition, each image is composed of its own set of rules; the ruleset that defines the instantiation of one image (or object) is different from any other. As a result, the ``rules'' of the whole canvas are inconsistent because it is a sum of multiple, externally-defined images. Temporally, the evolution of the canvas lends itself to different types of spatially local rulesets that are being implemented. At the beginning of the canvas, for example, users may be placing tiles randomly, which can later be used as ``seeds'' to scaffold an externally-existing image. During later stages, present images/objects are well-defined and pixel placement may be less random. Thus, the beginning and end stages of the canvas evolution maybe be more predictable given the state of the canvas than in the middle stage.

To explore whether this framework lends itself to the predictability of future canvas states, we trained a convolutional neural net (CNN) on certain temporal subsets of the canvas evolution and test its ability to predict future canvas states. We then compare its ability to make predictions with the results of the rule-based approach. Our reasoning is that if dynamics that emerge without external coordination from social constructs \textit{should} be derived simply from the canvas evolution, where objects that are a result of coordinated social efforts in subreddits \textit{can't}. The CNN model is used to demonstrate different areas of space and time over the canvas that can and can't be predicted by training on previous canvas frames.

We recognize that several extrinsic collaborations -- by which we mean interactions having taken place between users outside of the direct activity of editing the canvas -- contributed to the state of the canvas, which cannot be predicted from the state-evolution of the canvas alone. \cite{litherland2021} discusses the close relationship between objects that emerge and activity within some subreddits. Due to this, we cannot predict the emergence of objects that form due to collaboration and planning that occurred within subreddits, unless such information is fed into our training, including both social interactions (e.g. private discussion among factions of users outside the canvas painting activity) and cultural objects (a database of relevant images, flags, and logos, ideally including some attached cultural semantics). Here, we focus on collaboration ``rules'' that are purely spatial-based. In other words, this analysis tries to estimate ``rules'' that users use to decide which pixel to place where based on the current state of the canvas. Since both types of rules contribute to the state-evolution of the canvas, we acknowledge that our current approach is limited to only spatial state-based rules. In the future, we could use data from subreddits that hosted the external collaboration for the emergence of objects to gain a more complete predictive analysis.

In summary, our approach is aimed at addressing the following questions: What are the conditions that lead to an emergent structure? Are some structures emergent based on pixel-by-pixel interactions without social coordination? Finally, how do these objects differ from objects that are a direct result of social coordination? 

\section{Statistical rule-based approach}

We rendered snapshots of the canvas in 10-minute intervals to create a coarse-grained time-evolution of the canvas. This resulted in 682 total snapshots of the canvas over the whole dataset. In the times between each snapshot, we also counted the number of pixels placed and the number of unique users who placed pixels. What ``rules'' do users use to place a new pixel on the canvas, given the current state of the canvas? If the dynamics of the canvas state are open-ended, then these rules could change over time, possibly as a function of the state of the canvas \citep{e19090461}. Regardless if this evolution fits the requirements of open-endedness described by \cite{dolsonoee}, \cite{ttoee}, and \cite{bernatoee}, we leave open the possibility that the update rules from one frame to the other could change over time.

Artificial life's approach advocates for the understanding of complex systems from the bottom-up, by studying emergent properties in a generative way \citep{bedau2000open, frans2021questions}. The analysis may profit in from starting with the restricted frame of discrete dynamical systems such as cellular automata \citep{beer2014cognitive}, which give us access to a large range of tools from computer science and complex systems, such as counting the number of ways a finite region may transform from a defined region to another \citep{biehl2021investigating}. If we assume the canvas snapshots behave similarly to a synchronous 2-dimensional cellular automata (2D CA) with local interaction rules, then a static object that persists throughout snapshots are locally a static, unchanging set of rules. Here, we use a slightly different interpretation for ``rule'' than for 2D CA. For cellular automata, a rule is the set of single-pixel outcomes for all possible neighborhood states. Here, we take a more fine-grained definition of rule and assume a rule is a \textit{single} neighborhood state and an outcome for the center pixel. Using this interpretation, a cellular automata rule is a set of ``rules'' using our definition.

The Mona Lisa, for example, can be mapped to an area of pixels, each in a single state from a set of discreet colors. As a result, it is defined as a static set of rules that is only applied to the area that spans the image. The set of rules can have contradictory outcomes for the same neighborhood state, depending on how a neighborhood is defined. Over time, the pixels within the image do not change since it is a static image, and any pixel that is perturbed would be changed back according to the set of rules that define the static image, according to the image's set of rules. Our goal is to estimate the rule set for each image (and the canvas overall) and understand how these rule sets change over time as images appear, disappear, and compete for limited space in r/place.

We understand that it is possible (and likely) that rulesets that define images could include rules based on non-local neighborhood states. In addition, individual rules are likely to have different neighborhood sizes depending on the outcome pixel location in relation to the whole image. For example, a corner pixel of the Mona Lisa does not care about its immediate neighboring pixels outside of the image boundary. But the pixels in the center of the image do depend on the state of all its immediate neighbors, because the neighbors are within the boundary of the image. Images that are not as well-defined, such as the spreading ``Blue Corner'' and spreading black ``Void,'' have a simpler rule set: Simply to change any neighboring pixel to the image boundary blue or black, respectively. Different images and objects in the canvas will have different sets of rules that are intrinsic to the object type. But no matter the object, we argue that \textit{in order for an object to persist, the set of rules that define it must persist over time in that local space.}

\begin{figure}[h!]
  \centering
  \includegraphics[width=0.4\textwidth]{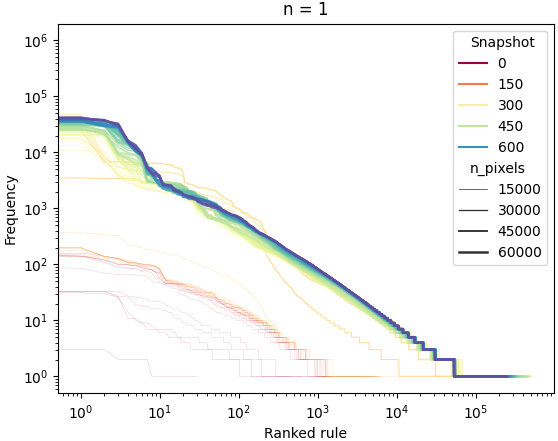}
    \caption{Rank-order frequency distributions of nearest-neighbor rules (radius = 1) between snapshots of the canvas using 10-minute intervals. Each line is the distribution of rules within the 10-minute time frame between snapshot.}
    \label{fig:rule_freq}
\end{figure}

\begin{figure}[ht!]
  \centering
  \includegraphics[width=0.4\textwidth]{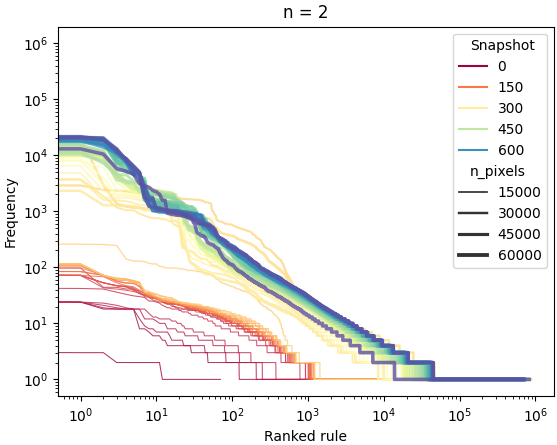}
    \caption{Rank-order frequency distributions of nearest-neighbor rules (radius = 2) between snapshots of the canvas using 10-minute intervals. Each line is the distribution of rules within the 10-minute time frame between snapshot.}
    \label{fig:rule_freq_2}
\end{figure}

Given 10-minute interval snapshots of the canvas, we map whatever local and non-local rules between snapshots as a 2D nearest-neighbor CA, with a nearest-neighbor radius of 1 and 2. Given the current color of a pixel and its 8 ($n=1$) or 24 ($n=2$) nearest neighbors, what color will the pixel be at the next time step? Regardless of the actual set of rules that govern the transition between canvas snapshots, even if these rules are changing over time, we encode all rules from one snapshot to the other using the 2D CA rule space. As a result, whichever rules are driving the evolution of the canvas, we understand that they are mapped into 2D CA rule spaces. Rules along the edges are excluded. Figures \ref{fig:rule_freq} and \ref{fig:rule_freq_2} show the rank-order frequency distributions of these rules over time, encoded in 2D CA rule spaces. Each line represents the rank-order distribution of rules between a snapshot and the snapshot at the next 10-minute interval. The line thickness represents the number of pixels placed during that time frame. If the thickness denoted the number of unique users that placed pixels instead, the results are visually identical in these plots.


\begin{figure}[h!]
  \centering
  \includegraphics[width=0.35\textwidth]{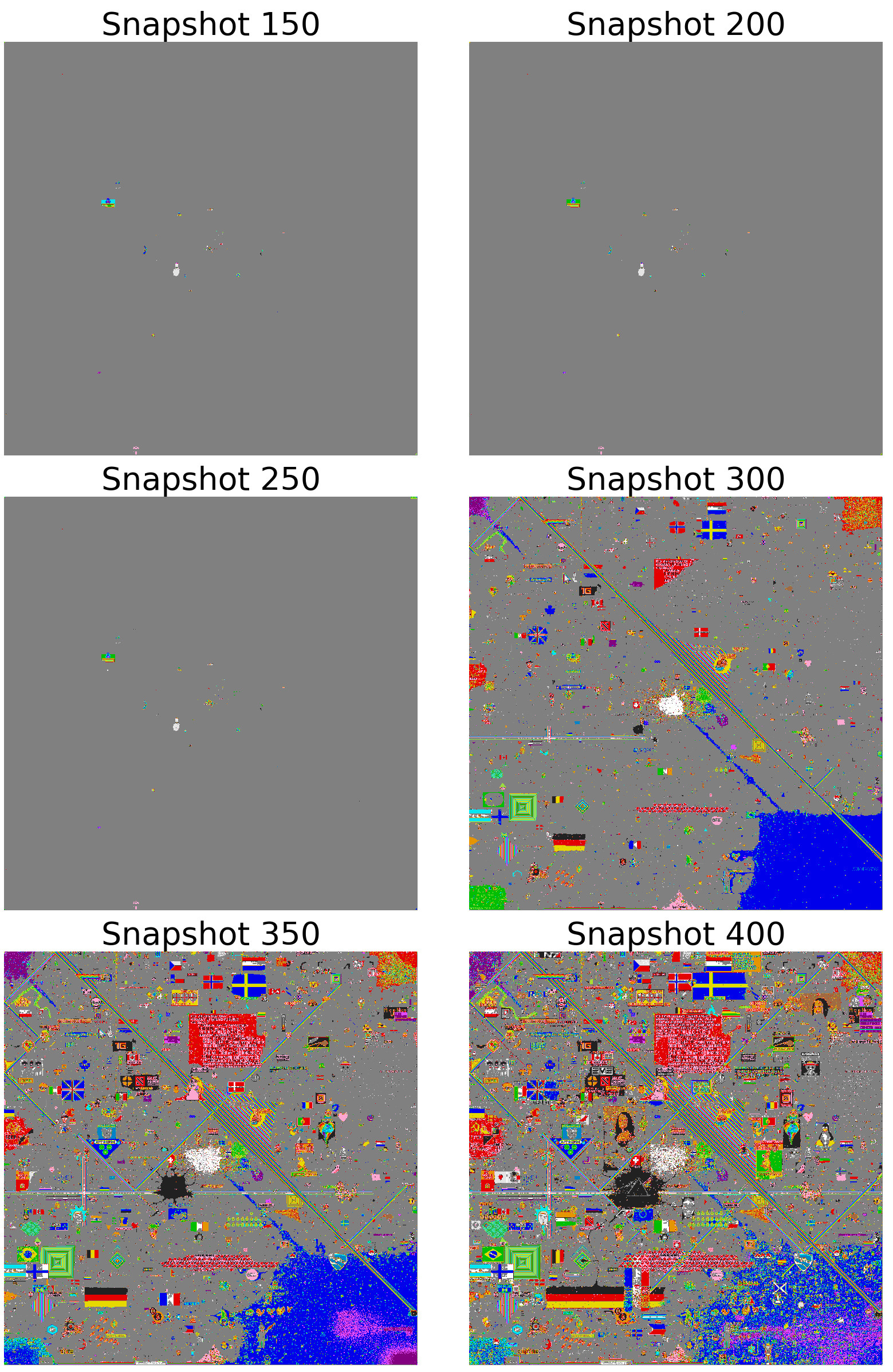}
    \caption{Key snapshots to illustrate the transition between sparse pixel-placement (before the 72-hour period) and dense pixel placement (the first few hours of the 72-hour period).}
    \label{fig:snapshots}
\end{figure}

To better understand the relationship between these rule frequency distributions and the objects present in the canvas, we have also included canvas snapshots in Figure \ref{fig:snapshots}. These snapshots correspond to the same time period in Figures \ref{fig:rule_freq} and \ref{fig:rule_freq_2} that spans the two ``groups'' of lines (red-orange to yellow-blue). 

\section{CNN-based neural network approach}

Next, we let a neural network learn the generative rules in order to generate the complex outcomes of the art canvas. This approach is similar to some works on growing neural cellular automata \citep{mordvintsev2020}, but with the following design changes. To learn the set of rules, we have trained a convolutional neural network (CNN) which, given a 3$\times$3-pixel kernel of a canvas snapshot in time, outputs the value of the central pixel of one of the successive frames. Running this neural network over all 3$\times$3 fragments for one frame of the 1000$\times$1000-pixel canvas, we end up with a 998x998-pixel canvas that may then be compared to the original art output in the r/place experiment. Similarly to the convolutional part of \cite{mordvintsev2020}, we have divided each of the RGB colors for the training in order to (1) simplify the learning process of the neural network, and (2) use the SSIM method (explained later) to measure the performance of the model.

\begin{figure}[ht!]
    \centering
    \includegraphics[width=0.45\textwidth]{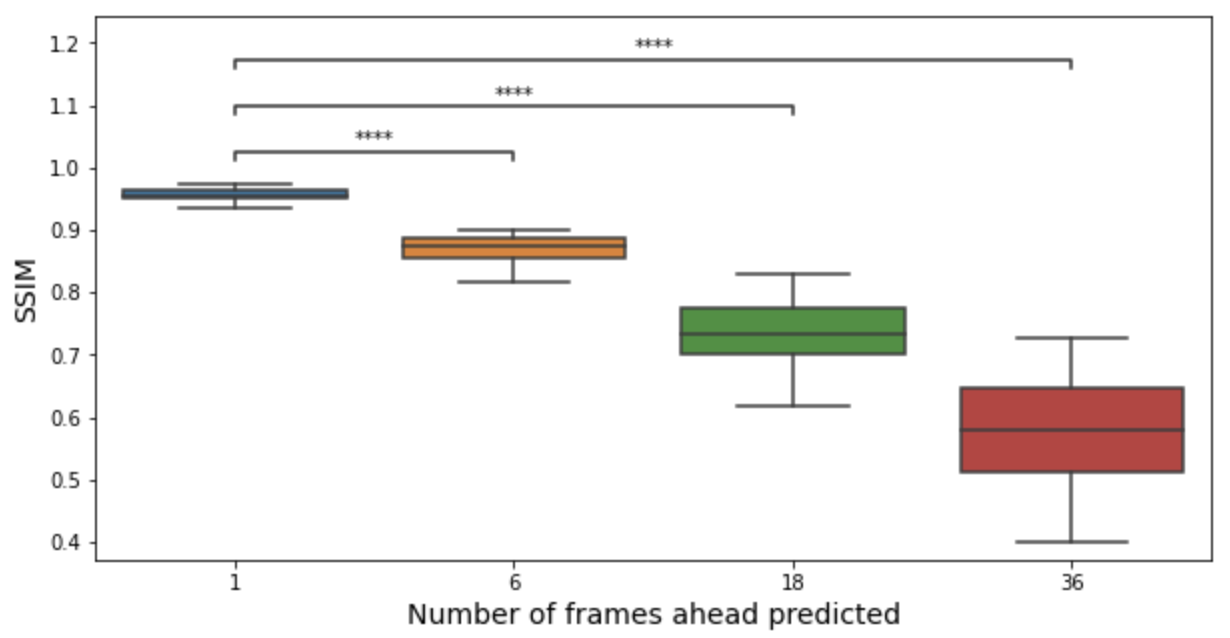}
    \caption{Model performance when predicting the following frame (10 minutes), the \nth{6} frame (1 hour), \nth{18} frame (3 hours), and \nth{36} frame (6 hours).}
    \label{fig:nnFrameAhead}
\end{figure}

\begin{figure*}[h!]
    \centering
    \includegraphics[width=0.9\textwidth]{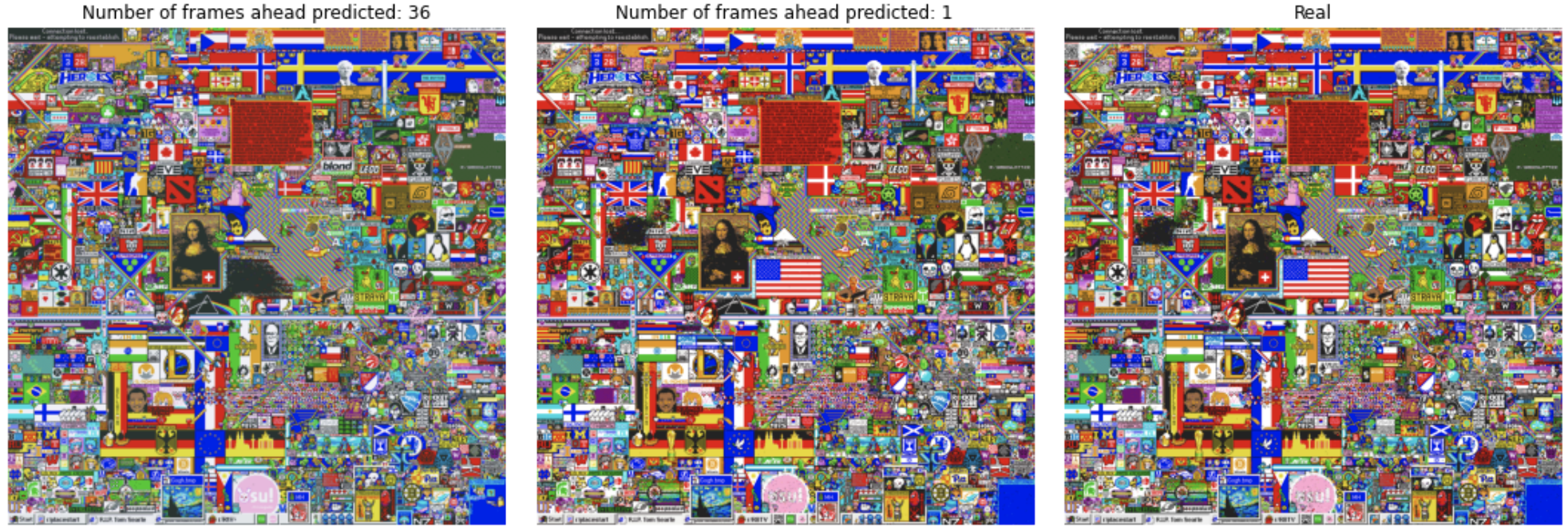}
    \caption{\textit{Left:} Prediction by the model for the last frame when predicting 36 frames ahead. \textit{Center:} Same prediction but one time frame ahead. \textit{Right:} Actual frame of the social experiment on r/place. Whereas the output of the model when predicting one time frame ahead is nearly the same as the final one (all the objects appear and the average SSIM for the RGB colors is 0.95), there are several objects that do not appear in the output image when predicting 36 time frames ahead. Some examples are the dove on the EU flag, the objects within the blue corner, or the USA flag in the center part of the image. These examples are all extrinsic collaborations between users, which cannot be predicted unless the model is trained with extra inputs.}
    \label{fig:nnExampleImage}
\end{figure*}

To deepen the analysis, we have trained the neural network under different conditions. We have first trained the model to predict not only the next frame, but also the \nth{6}, \nth{18}, and \nth{36} frame. Considering that the time gap between frames is 10 minutes, these frame gaps correspond to 10 minutes, 1 hour, 3 hours, and 6 hours, respectively. Figure \ref{fig:nnFrameAhead} shows the results of the model for this predictive task. The results indicate that the further the frame to be predicted, the less significantly similar the frame created by the neural network is compared to the ground truth. Figure \ref{fig:nnExampleImage} shows two different outputs of the model for the last frame compared to the last frame of the original social experiment. 

\begin{figure}[h!]
    \centering
    \includegraphics[width=0.45\textwidth]{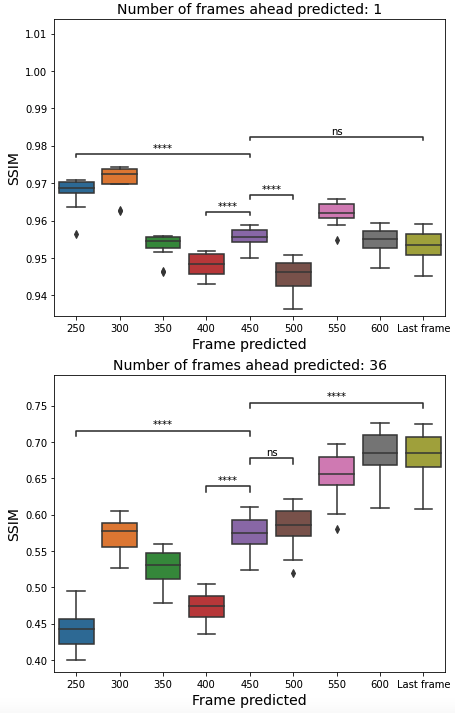}
    \caption{Model performance when predicting different frames of the video. On the top image, results when predicting one time frame ahead. On the bottom image, results when predicting 36 time frames ahead.}
    \label{fig:nnFramePredicted}
\end{figure}

We have also compared the prediction of the model in different timestamps of the video, assessing the model performance every 50 frames ($\sim$ 8 hours). When predicting one frame ahead, we observed that the accuracy changed over time as it significantly decreased for the \nth{400} and \nth{500} frame when compared to the \nth{450} frame (p-value $<$ 0.001), and significantly increased for the \nth{250} and \nth{300} frame (p-value $<$ 0.001). Instead, when predicting 36 frames ahead, the model became significantly better at predicting the last frames compared to the first frames of the video. Compared to the prediction of the \nth{400} frame, the prediction for the \nth{250} was significantly lower (p-value $<$ 0.001) while the prediction of the last frame was significantly higher (p-value $<$ 0.001). All these results are displayed in Figure \ref{fig:nnFramePredicted}.

\section{Discussion}


Results from the statistical rule-based approach indicate that rules for the overall canvas encoded in the $n=1$ 2D CA space are more frequently repeated than for $n=2$. However, this could be due to the fact that the size of the rule space is considerably smaller for $n=1$ than it is for $n=2$. In the future, we plan on comparing rule frequencies based on canvas snapshots with rule frequencies between white noise images. In the case of white noise images, the rule space would be sampled at random. But since the size of the images are only 1000$\times$1000 pixels, the frequency distributions are limited to the size of the canvas (not all rules can be expressed at once due to the size). This analysis would provide a baseline approximation for how much of the rule space is being explored in the canvas as compared to a random sampling of the rule space.

We are interested in which rules are expressed the most frequently over time. Are the same rules the most popular over the whole evolution, or are do the most popular rules change over time? Furthermore, are these distributions an indication of the number of well-formed objects present in the canvas at that time?

As these canvas update rules change over time (as indicated in Figures \ref{fig:rule_freq} and \ref{fig:rule_freq_2}), we also want to better understand the relationship between the state of the canvas and which rules are expressed the most. This approach explicitly linked the state of the canvas to update rules, since update rules were derived from canvas states by encoding the rules in 2D CA space. Which rule space encodings are the most deterministic and how can we determine that from states of the canvas? It is possible that the most deterministic and predictive rule encodings change over time as well. A rule encoding such as 2D CA $n=1$ is best for some points in time and 2D CA $n=2$ for other points in time. But this may also change over the space of the canvas as well. The ``Blue Corner'' (seen in snapshots 300-400 in Figure \ref{fig:snapshots}) may use a set of rules based on nearest-neighbor states, but the Mona Lisa appearing in snapshot 400 may use a much more constrained rule set based on an external, well-defined pattern mapped to a set of pixels. Future analyses could compare these rule frequency distributions with distributions based on random neighborhoods, including non-local neighborhoods with non-adjacent pixels.

The results from the CNN-based neural network approach indicate that the collaboration rules become easier to infer as we decrease the time gap between the input frame and desired frame to reconstruct. Perhaps more surprisingly, the behavioral rules followed by the users to modify the art piece were not kept constant over time, being statistically more difficult to infer for the middle time frame of the experiment, in spite of edits and interactions remaining intense until the end of the experiment \citep{howWe2017}. This implies that the user decisions to modify a pixel are more based on short-term rather than long-term contexts in time, while being also influenced by the number and type of objects contained in the image, as they increase over time until the last hours where the canvas becomes more static. Even though there were ``outside'' collaborations between users that cannot be predicted from the state-evolution of the canvas, these findings indicate that there were general trends over the art piece that were found by the model. Hence, it was possible to get some predictability over certain objects of the canvas, such as the ones growing linearly over time such as the blue corner or the German flags. One example is the output image of the model shown in Figure \ref{fig:nnExampleImage}, where local dependencies were predicted by the model while global dependencies, such as the dove on the EU flag, were not inferred.

Nevertheless, this predictability would probably increase if analyzing each subset of users that were collaborating together to create new objects. So, we leave to future work the implementation of this same model for sets of users to gain a more complete understanding of the decisions taken in relation with complex interactions.

\section{Conclusion}

This work focused on the study of the social dynamics of interacting agents who are able to collectively create an art piece such as the r/place. To gain a better understanding of these dynamics, we use a statistical 2D cellular-automata rule-based approach to address spatially local rules that don't change over time as a rough estimation of the number of objects present in the canvas during that snapshot and a CNN to evaluate and compare the predictability of future states based on prior canvas evolution.

For the rule-based approach, we mapped the rules between snapshots using a 2D nearest-neighbor CA with a radius of 1 and 2 and have found that rules that rules encoded in the $n=1$ 2D CA space are more frequently repeated than for $n=2$. In addition, the number of rules expressed increases over time, which reflects the number of objects present in the canvas over various snapshots. After $t=300$, the distribution remains somewhat linear in log-log space, particularly for the $n=1$ encoding. The $n=2$ encoding has more diverse rank-order rule distributions over time, which suggests that additional rule encodings could be explored. For example, objects such as ``The Blue Corner'' and ``The Void'' are defined by changing any adjacent pixel blue or black, respectively, which is a particular subset of all possible 2D CA rules. On the other hand, the Mona Lisa image is a set of rules that are dependent on relative position to each other. In the future, we work towards generalizing this approach to exploring a wider variety of rule encodings and finding a meaningful way to normalize and compare results across them.

Regardless of whatever rule encoding is best used to describe the dynamics of r/place, we let a neural network learn a set of generative rules to generate complex outcomes using a window of 3$\times$3. Our results show that the collaboration rules become easier to predict as we decrease the time gap between the input and output frame, as well as the behavioral rules changed over time. More predictability is observed for shorter time frames and when the board displays less complexity and towards the beginning, but also surprisingly towards the end as well, while participation and interactions were still increasing. This suggests that rules that define the emergence of objects are often the result of external social collaborations in subreddits. However, some of the canvas behavior can be predicted from pixel-by-pixel dynamics alone. This suggests that some canvas behavior can be determined from rule-based interactions alone, whatever those rules may be.

Overall, our results may bring about many interesting questions about the dynamics of the art piece and social cooperation within a digital medium. We think that further work could profit from focusing on finding which rule encodings are more deterministic. For example, one may implement these same methods for each set of users, to then compare their outcomes and inferred rules to make sense of local and global dynamics in social behavior. Having combined the insights of a rule-based approach with a machine learning one, we were able to discover diverse aspects of the social dynamics from a promising dataset. This seems to constitute a clear indication that both tools may be advantageously combined for this field of research.

\footnotesize
\bibliographystyle{apalike}
\bibliography{library} 

\end{document}